\documentclass[aps,pra,twocolumn,groupedaddress]{revtex4-1}
\usepackage{dcolumn}
\usepackage{graphicx}
\usepackage{amsmath}
\usepackage{amsfonts}
\usepackage{color}

\begin{document}


\title{Quantum entanglement in the spatial symmetry breaking phase transition  \\
 of a driven-dissipative Bose-Hubbard dimer}


\author{Wim Casteels}
\author{Cristiano Ciuti}
\affiliation{ Laboratoire Mat\'eriaux et Ph\'enom\`enes Quantiques, Universit\'e Paris Diderot, CNRS UMR 7162, Sorbonne Paris Cit\'e, 10 rue Alice Domon et Leonie Duquet 75013 Paris, France}


\date{\today}

\begin{abstract}
We theoretically explore quantum correlation properties of a dissipative Bose-Hubbard dimer in presence of a coherent drive. In particular, we focus on the regime where the semiclassical theory predicts a bifurcation with a spontaneous spatial symmetry breaking.
The critical behavior in a well defined thermodynamical limit of large excitation numbers is considered and analyzed within a Gaussian approach. The case of a finite boson density is also examined by numerically integrating the Lindblad master equation for the density matrix. We predict the critical behavior around the bifurcation points accompanied with large quantum correlations of the mixed steady-state, in particular exhibiting a peak in the logarithmic entanglement negativity.
\end{abstract}

\pacs{}

\maketitle

\section{Introduction}
In recent years there has been a growing interest in driven-dissipative photonic systems for the realization of correlated quantum states (see for example Refs. \cite{book:111220,RevModPhys.85.299, ANDP:ANDP201200261,2016arXiv160500383H, 2016arXiv160404433N} for recent comprehensive reviews). One direction of research of particular relevance is the study of dissipative phase transitions and quantum criticality in these out-of equilibrium-systems \cite{PhysRevLett.110.257204,PhysRevA.86.012116,PhysRevB.82.100507, PhysRevA.88.053627, PhysRevX.4.031043, Hamel:2015aa, PhysRevLett.116.070407, PhysRevX.5.031002, PhysRevB.93.014307, PhysRevA.93.023821,2016arXiv160206553J}. In particular a recent study by Carmichael has emphasized that critical phenomena can emerge in the thermodynamical limit defined by a large number of photons even with no spatial degrees of freedom (only one cavity) \cite{PhysRevX.5.031028}. The simplest system where there can be an interplay between spatial degrees of freedom and interactions is the two-cavity dimer. The driven-dissipative Bose-Hubbard dimer has been the subject of recent investigations and can be realized, e.g., in a system of two coupled nonlinear optical cavities (see the inset of Fig. \ref{Fig1} for a sketch). Already at the semiclassical level a very rich behavior is predicted with parametric instabilities \cite{PhysRevB.77.125324} and a spontaneous symmetry breaking \cite{Maes:06, Rodrigues2013}. 

Recently, an experimental set-up with coupled photonic-crystal lasers nicely revealed this spontaneous symmetry breaking \cite{Hamel:2015aa}. The driven-dissipative Bose-Hubbard dimer has also been realized with various other experimental platforms such as semiconductor microcavities \cite{Abbarchi:2013aa,PhysRevLett.108.126403, PhysRevLett.105.120403,2016arXiv160207114R} and superconducting circuits \cite{PhysRevLett.113.110502,PhysRevX.4.031043}. 
A Bose-Hubbard dimer has also been studied for the realization of unconventional photon blockade effects in systems with weak nonlinearity \cite{PhysRevLett.104.183601,PhysRevA.83.021802}.
The closed system analog is the bosonic Josephson junction for which the semiclassical approach also predicts a symmetry breaking \cite{PhysRevA.66.013602, PhysRevE.74.056608}, observed experimentally with ultracold gases \cite{PhysRevLett.105.204101}. For this system various theoretical efforts have been devoted to providing a quantum description that goes beyond the semiclassical approximation (see for example Refs. \cite{PhysRevA.88.033601, PhysRevA.88.033608, PhysRevA.88.063606, PhysRevA.85.043625}).  
  
In this paper, we explore the  physics of the driven-dissipative Bose-Hubbard dimer in the region where the semiclassical approach predicts two bifurcation points and a spontaneous spatial symmetry breaking.  An analytical quantum description of the critical behavior  is provided by considering the Gaussian fluctuations around the semiclassical result. This gives an exact description in a thermodynamical limit of large excitation numbers, which is carefully defined. In particular, we determine the critical behavior of the von Neumann entropy and of the logarithmic negativity, which is a measure of entanglement for mixed states. The finite-size behavior (finite number of bosons) is examined by numerically integrating the master equation for the density-matrix. 

In section \ref{Sec2} the driven-dissipative Bose-Hubbard model is introduced and the semiclassical prediction of spontaneous symmetry breaking is discussed. Furthermore, in Section \ref{Sec2} the concept of a well-defined thermodynamic limit with an infinite number of photons is introduced. In section \ref{Sec3} the role of the quantum fluctuations around the semiclassical prediction are examined up to quadratic order analytically and compared with numerical simulations which reveals the presence of a quantum critical region for a finite photon number. Then, in section \ref{Sec4}, the behavior of the von Neumann entropy and the logarithmic negativity are examined.
Finally, in section \ref{Sec5}, the conclusions and perspectives are presented.

\section{Semiclassical prediction of spontaneous symmetry breaking \label{Sec2}}
We start by considering the following Hamiltonian (with $\hbar = 1$):
\begin{equation}
\hat{H}_{BH} = -J\left(\hat{a}_1^\dagger\hat{a}_2+\hat{a}_2^\dagger\hat{a}_1\right) +  \sum_{j=1,2}\left(\omega_c\hat{a}_j^\dagger\hat{a}_j+U\hat{a}_j^\dagger\hat{a}_j^\dagger\hat{a}_j\hat{a}_j\right),
\label{Eq: SysHam}
\end{equation} 
where  $\hat{a}^{\dagger}_j$ ($\hat{a}_j$) is the creation (destruction) operator of a boson on site $j \in \{1,2 \}$.
The first term represents the hopping between the two sites with rate $J$. The second term describes the energy of the boson modes whose frequency is $\omega_c$. The boson-boson interaction is quantified by the on-site energy $U$ (we will consider a repulsive interaction with $U > 0$).  The corresponding linear system ($U = 0$) consists of two normal modes typically denoted as the bonding ($+$) and the anti-bonding ($-$) modes. The corresponding bosonic  operators are $\hat{a}_{\pm} = (\hat{a}_1 \pm \hat{a}_2)/\sqrt{2}$ and the corresponding mode frequencies are $\omega_\pm = \omega_0 \mp J$. 

The dimer is considered to be driven coherently with an amplitude $F_i$ on site $i$ and frequency $\omega_p$,  which is the same on both sites. This is described by adding the following drive term to the Hamiltonian:  
\begin{equation}
\hat{H}_{p}(t) = \sum_{j=1,2}\left(F_j e^{-i\omega_pt}\hat{a}_j^\dagger+F^*_j e^{i\omega_pt}\hat{a}_j\right).
\end{equation} 
The total system Hamiltonian is $\hat{H} = \hat{H}_{BH} + \hat{H}_{p}$. In the quantum optical context, such Hamiltonian can be implemented by two coupled cavity resonators with a Kerr photon-photon nonlinearity.
For sake of simplicity we will work in the frame rotating at the drive frequency $\omega_p$, which eliminates the time dependence of the Hamiltonian. The relevant parameter is the  detuning $\Delta = \omega_p - \omega_c$.
The boson losses are described perturbatively within the Born-Markov approximation resulting in the following Lindblad-master equation for the dimer reduced density matrix $\hat{\rho}$
\begin{equation}
i\frac{\partial\hat{\rho}}{\partial t}=\left[\hat{H},\hat{\rho}\right] + i\frac{\gamma}{2}\sum_{j=1,2}\left[2\hat{a}_j\hat{\rho}\hat{a}^\dagger_j - \hat{a}_j^\dagger\hat{a}_j\hat{\rho}-\hat{\rho}\hat{a}_j^\dagger\hat{a}_j \right],
\label{eq:Master}
\end{equation}
where $\gamma$ is the loss rate. 

\begin{figure}[h]
  \includegraphics[scale=0.6]{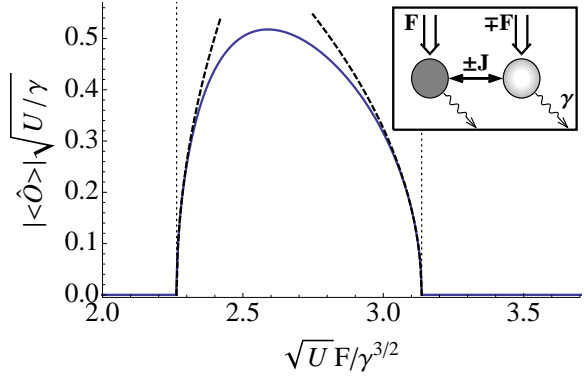}
  \caption{\label{Fig1} Semiclassical mean-field results describing the spatial symmetry breaking in the driven-dissipative Bose-Hubbard dimer. The rescaled order parameter $|\langle \hat{O} \rangle |\sqrt{U/\gamma}$, with $|\langle \hat{O} \rangle | =  |\langle \hat{a}_1 + \hat{a}_2  \rangle |$, is depicted (solid line) as a function of the dimensionless quantity $\sqrt{U}F/ \gamma^{3/2}$. These results become exact in the thermodynamic limit with an infinite number of photons, obtained for $\sqrt{U}F$ fixed and $F \rightarrow + \infty$. There are two bifurcation points (indicated by the dotted lines): in-between the symmetry is broken (non-zero order parameter $|\langle \hat{O} \rangle |$). The dashed lines are fits  with a square root dependence in the symmetry broken phase around the two bifurcation points. Other parameters: $J = 2.5\gamma$ and $\Delta = -1.5\gamma$. The inset gives a  sketch of the driven-dissipative Bose-Hubbard dimer and the two possible equivalent configurations: the first one is characterized by a positive hopping ($J > 0$) and a $\pi$ phase difference between the driving fields; the second equivalent configuration has a negative hopping strength ($J < 0$) and the same driving phase on the two sites. }
\end{figure}

The semiclassical approach is achieved by replacing the operators $\hat{a}_i$ with complex amplitudes $\alpha_i = \langle a_i\rangle$ satisfying the following nonlinear equations:
\begin{equation}
\label{GPEq}
\begin{cases}
    i\frac{\partial \alpha_1}{\partial t} = \left(-\Delta - i \frac{\gamma}{2}+ 2U \left| \alpha_1\right|^2 \right)\alpha_1 - J\alpha_2 + F_1 = 0,\\
    i\frac{\partial \alpha_2}{\partial t} =\left(-\Delta - i \frac{\gamma}{2}+ 2U \left| \alpha_2\right|^2 \right)\alpha_2 - J\alpha_1 + F_2 = 0.
	\end{cases}
\end{equation}   
From now on we will consider a spatial driving configuration that excites selectively  the anti-bonding mode, i.e. $J>0$ and $F = F_1 = - F_2$ (see inset of Fig. \ref{Fig1}). In this case the Lindblad-master equation (\ref{eq:Master}) has a discrete $\mathbb{Z}_2$ symmetry described by the transformation $\hat{a}_1 \leftrightarrow -\hat{a}_2$. At the semiclassical level this corresponds to the symmetry $\alpha_1 \leftrightarrow -\alpha_2$ of the Eqs. (\ref{GPEq}). We consider this particular driving scheme to obtain a regime with only two stable solutions that exhibit a spontaneous symmetry breaking. An alternative possibility to obtain such a regime is a negative hopping parameter $J <0$ and $F = F_1 =  F_2$ (see inset of Fig. \ref{Fig1}), as was the case in the experimental set-up presented in Ref. \cite{Hamel:2015aa}.    

The mean-field Eqs. (\ref{GPEq}) become exact in the limit of an infinite number of photons.  A simple scaling analysis of Eqs. (\ref{GPEq}) reveals that taking the limit $F \rightarrow + \infty$, while keeping the product $\sqrt{U}F$ fixed, results in an infinite number of photons ($|\alpha_i|^2 \rightarrow + \infty$ with $i \in \{1,2\}$) and a well defined thermodynamical limit. This can be seen by substituting $\alpha_i' = \sqrt{U}\alpha_i$ in Eqs. (\ref{GPEq}) which makes the rescaled equations  depend only on $\sqrt{U}F$, $\Delta$, $J$ and $\gamma$. Keeping $\sqrt{U}F$ fixed, we clearly have that   $|\alpha_i|^2 = |\alpha_i'|^2/U \to + \infty$  in the limit $U \rightarrow 0$, hence this is the thermodynamical limit in our problem. Therefore we will present the results as a function of the dimensionless quantity $\sqrt{U}F/\gamma^{3/2}$.

For the considered  spatial driving configuration and a frequency below the anti-bonding resonance and sufficiently blue detuned with respect to the bonding mode resonance
($\omega_+ + \sqrt{3}\gamma/2 < \omega_p < \omega_- $)  there is a range of values for the drive amplitude $F$ for which the Eqs. (\ref{GPEq}) admit three solutions. Two of these are dynamically stable and exhibit a spontaneous symmetry beaking. To examine this further, we introduce the operator $\hat{O} = \hat{a}_1 + \hat{a}_2$. Note that the expectation value of $\hat{O}$ is zero for states which are symmetrical with respect to the transformation $\hat{a}_1  \leftrightarrow -\hat{a}_2$ of the Lindblad-master equation (\ref{eq:Master}). This shows that the expectation value of $\hat{O}$ can be used as an order parameter signaling a spatial symmetry breaking. In Fig. \ref{Fig1} the semiclassical prediction for $|\langle\hat{O}\rangle|$ is presented as a function of $\sqrt{U}F/\gamma^{3/2}$ for a detuning $\Delta = -1.5\gamma$ and hopping strength $J = 2.5 \gamma$. In particular, the normalized quantity $|\langle\hat{O}\rangle| \sqrt{U/\gamma}$ is considered because due to the scaling properties of the mean-field equations the corresponding universal behavior does not depend on $U$ alone, but only on $\sqrt{U}F/\gamma^{3/2}$, which is well defined in the considered thermodynamical limit. The calculations show two bifurcation points between which the symmetry is broken. Close to the bifurcation points, in the symmetry-broken phase the order parameter has the power-law behavior $|\langle \hat{O} \rangle |  \sqrt{U/\gamma} \propto (\sqrt{U}F - A_c)^{1/2}$ (see dashed lines in Fig \ref{Fig1}), where $A_c$ is a constant depending on the detuning $\Delta/\gamma$. In other words, there is a critical exponent $1/2$ for the order parameter. The two stable symmetry breaking solutions of Eqs. (\ref{GPEq}) have the same value of $|\langle\hat{O}\rangle|$ and are related by the transformation $\{\alpha_1,\alpha_2\} \leftrightarrow -\{\alpha_2,\alpha_1\}$ . We emphasize that the results presented  in the following are the same for the two symmetry-breaking solutions. 

\section{Quadratic fluctuations and numerical simulations \label{Sec3}}
Systematic corrections to the semiclassical predictions can be obtained through an expansion of the fields around the mean-field amplitudes. Truncating this expansion at the quadratic order allows to solve the resulting equations of motion for the system exactly. We express the fields as $\hat{a}_i = \alpha_i + \hat{\delta}_i$, where $\hat{\delta}_i$ is the operator describing the correction to the mean-field result. In general a Gaussian system is completely described by the covariance matrix which contains the quadratic correlation functions (see for example Ref. \cite{1751-8121-40-28-S01}). We note that for a photonic system all the considered quadratic correlation functions are experimentally accessible, e.g., through a homodyne detection scheme \cite{1464-4266-7-12-021, Boulier:2014aa}. The equations of motion for these quadratic correlation functions form a closed set which for the considered case of two coupled modes corresponds to $6$ linear independent equations. This is in stark contrast to the general case of a nonlinear system with an infinite hierarchy of coupled equations of motions for the correlation functions at all orders \cite{2016arXiv160500882C}. Within the Gaussian truncation, the equations of motion for the local quadratic correlation functions for site $1$ are:
  \begin{eqnarray}
    i\partial_t \langle \hat{\delta}_1^\dagger\hat{\delta}_1\rangle = &&-i\gamma \langle \hat{\delta}_1^\dagger\hat{\delta}_1\rangle + 2U \alpha_1^2\langle \hat{\delta}_1^{\dagger2}\rangle - 2U \alpha_1^{*2}\langle \hat{\delta}_1^{2}\rangle \nonumber \\
    && - J\left(\langle \hat{\delta}_1^\dagger\hat{\delta}_2\rangle - \langle \hat{\delta}_2^\dagger\hat{\delta}_1\rangle \right); \\
    i\partial_t  \langle \hat{\delta}_1^{2}\rangle = && 2\left(\Delta + 4U\left| \alpha_1 \right|^2 - i\frac{\gamma}{2}\right) \langle \hat{\delta}_1^{2}\rangle \nonumber\\
     &&+ 2U\alpha_1^2 \left(1 + 2 \langle \hat{\delta}_1^\dagger\hat{\delta}_1\rangle\right) - 2 J \langle \hat{\delta}_1\hat{\delta}_2\rangle.
  \label{eq:Gauss}
\end{eqnarray}   
Similarly, the equations of motion for the local second order expectation values for site $2$ are obtained by the substitution $1 \leftrightarrow 2$. For the non-local expectation values we obtain: 
\begin{eqnarray}
    i\partial_t \langle \hat{\delta}_1^\dagger\hat{\delta}_2\rangle =&& \left(4U \left|\alpha_2\right|^2-4U \left|\alpha_1\right|^2-i\gamma \right)\langle \hat{\delta}_1^\dagger\hat{\delta}_2\rangle + 2U \alpha_2^2\langle \hat{\delta}_1^{\dagger}\hat{\delta}_2^{\dagger}\rangle \nonumber \\ 
    && - 2U \alpha_1^{*2}\langle \hat{\delta}_1\hat{\delta}_2\rangle - J\left(\langle \hat{\delta}_1^\dagger\hat{\delta}_1\rangle - \langle \hat{\delta}_2^\dagger\hat{\delta}_2\rangle \right) ,\\
 i\partial_t \langle \hat{\delta}_1\hat{\delta}_2\rangle =&& \left(2\Delta + 4U \left|\alpha_2\right|^2 + 4U \left|\alpha_1\right|^2-i\gamma \right)\langle \hat{\delta}_1\hat{\delta}_2\rangle \nonumber \\
 &&+ 2U \alpha_2^2\langle \hat{\delta}_1\hat{\delta}_2^{\dagger}\rangle + 2U \alpha_1^{2}\langle \hat{\delta}_1^\dagger\hat{\delta}_2\rangle - J\left(\langle \hat{\delta}_2^2\rangle + \langle \hat{\delta}_1^2\rangle \right) .
  \label{eq:Gauss}
\end{eqnarray}   
Note that as for the semiclassical Eqs. (\ref{GPEq}) the substitution $\alpha_i' = \sqrt{U}\alpha_i$ removes the explicit dependence on the nonlinearity $U$ for these equations of motion. This results in an effective description for the system dynamics with Gaussian states which becomes exact in the thermodynamic limit. From now on we will consider the steady-state of the system which corresponds to setting the time derivatives in the equations of motion equal to zero. 

\begin{figure}[h]
  \includegraphics[scale=0.65]{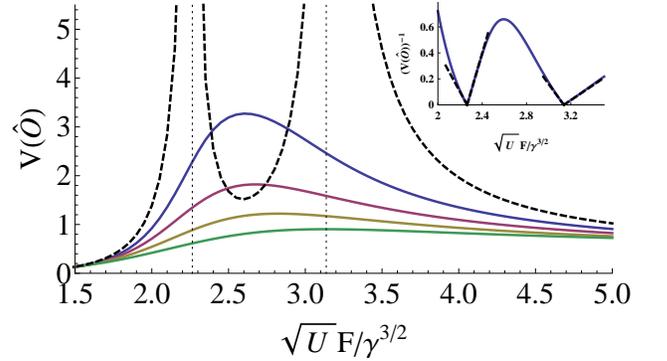}
  \caption{\label{Fig2} The variance of the order parameter $V(\hat{O})$ as a function of $\sqrt{U}F/\gamma^{3/2}$. The dashed curve is the Gaussian approach which is exact in the thermodynamic limit ($\sqrt{U} F$ fixed and $U \to 0$). Solid lines depict numerical solutions of the full master equation (\ref{eq:Master}) for  $U/\gamma = 0.1, 0.25, 0.5$ and $1$ (from top to bottom). Inset: the inverse of the variance $V\left(\hat{O}\right)^{-1}$ as a function of $\sqrt{U}F/\gamma^{3/2}$ in the thermodynamic limit. The dashed lines in the inset are fits with  $V\left(\hat{O}\right) \propto (\sqrt{U} F  - A_c)^{-1}$ around the bifurcation points, where $A_c$ is a constant. Other parameters: $J = 2.5\gamma$ and $\Delta = -1.5\gamma$. } 
\end{figure}   

In Fig. \ref{Fig2}, we start by looking at fluctuation properties.  In particular, we analyze the variance of the order parameter, namely $V(\hat{O}) =\langle |\hat{O}|^2 \rangle  - |\langle \hat{O} \rangle |^2 = \langle \hat{\delta}_1^\dagger\hat{\delta}_1\rangle + \langle \hat{\delta}_2^\dagger\hat{\delta}_2\rangle + 2Re[\langle \hat{\delta}_1^\dagger\hat{\delta}_2\rangle] - |\langle \hat{\delta}_1 + \hat{\delta}_2 \rangle|^2$ as a function of $\sqrt{U}F$ for the same parameters as in Fig. \ref{Fig1}. We show also the results obtained from a numerical integration of the full master equation (\ref{eq:Master}) with different values of the nonlinearity $U$. These results have been obtained by determining the steady-state solution of the master equation (\ref{eq:Master}) in the Fock number state basis of the fluctuation operators $\{\hat{\delta}_i\}$, with a maximal considered cutoff of 16 excitations per site. In particular, 
the steady-state solution is obtained by diagonalization of the Liouvillian linear superoperator associated to the master equation, where the steady-state solution corresponds to the zero eigenvalue. From the behavior far from the symmetry breaking region, we see that the numerical results indeed tend to the Gaussian description as the interaction strength $U$ is decreased, while keeping $\sqrt{U} F$ constant. In the thermodynamic limit the variance of the order parameter diverges at the bifurcation points, as expected from the general theory for phase transitions. This can be seen more clearly in the inset of Fig. \ref{Fig2} where the inverse of the variance is presented. In particular,  at the bifurcation points the divergence of the variance follows the power law $V(\hat{O}) \propto (\sqrt{U} F - A_c)^{-1}$, thus with a critical exponent $1$. We would like to stress again that the variance of the order parameter $V(\hat{O})$ is experimentally accessible through a homodyne quantum optical detection scheme \cite{1464-4266-7-12-021, Boulier:2014aa}. 

The numerical results with a finite photon density reveal the presence of a quantum critical region around the bifurcation points where the quantum fluctuations are not captured by the Gaussian approach. In this case the higher order terms beyond the quadratic approximations become relevant. 
We note that since the numerical integration scheme relies on a cutoff for the maximal number of photons it is unfortunately not possible to fully explore numerically the transition to the thermodynamic limit.      
Note that for $U \to 0$ the numerical calculation is most challenging at the bifurcation points, because the fluctuations diverge and one should have to take an arbitrary large cut-off. In the region between the two bifurcation points, convergence is much easier (the gaussian result is finite) and we have indeed verified (not shown) that the numerical results tend well to the gaussian theory.

\section{Von Neumann entropy and logarithmic entanglement negativity \label{Sec4}}
For a two-mode Gaussian system analytical expressions exist for many physical properties. We will consider two that play an important role in the theory of quantum information: the logarithmic entanglement negativity $E_N$ and the von Neumann entropy $S$. The entanglement negativity is defined as $ \mathcal{N}=  \sum_i \left(|\lambda_i| - \lambda_i\right)/2$, where the $\lambda_i$ are the eigenvalues of the operator which is obtained by performing a partial transposition of the density matrix $\hat{\rho}$ with respect to one of the subsystems (one of the two sites for the present dimer system). A finite value of the entanglement negativity $\mathcal{N}$ \cite{PhysRevA.65.032314} is a sufficient condition for bipartite entanglement and is used as a measure of it.  We will consider the closely related logarithmic negativity $E_N = \text{ln}(2\mathcal{N} + 1)$. The von Neumann entropy is defined as
$S =  -\text{Tr}[\hat{\rho}\ln[\hat{\rho}]]$ and measures the mixed character of the steady state.
In the case of a two-mode Gaussian system an exact analytical expression in terms of the second order expectation values has been derived in Ref. \cite{doi:10.1007/s11080-005-5730-2} for the entanglement negativity and in Ref.  \cite{0953-4075-37-2-L02} for the von Neumann entropy. 

\begin{figure}[h]
  \includegraphics[scale=0.65]{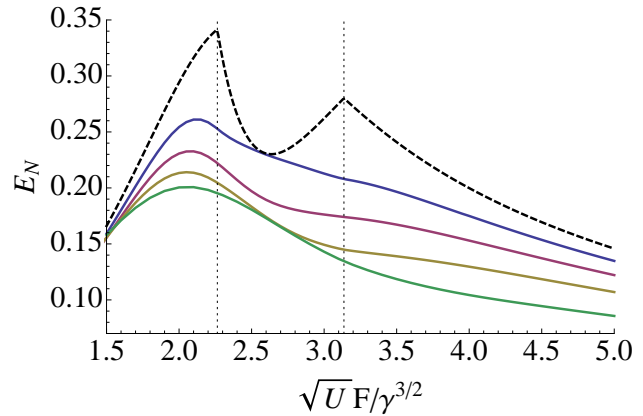}
  \caption{\label{Fig4} The logarithmic entanglement negativity $E_N$ as a function of $\sqrt{U}F/\gamma^{3/2}$. The dashed curve is the result for the Gaussian approach, which is valid in the thermodynamic limit ($U \rightarrow 0$ with $\sqrt{U} F$ fixed). The full curves correspond to finite values of the nonlinearity $U/\gamma = 0.1, 0.25, 0.5$ and $1$ (from top to bottom respectively). Other system parameters are the same as in Fig. \ref{Fig1}.}
\end{figure}

In Fig. \ref{Fig4} the logarithmic entanglement negativity $E_N$ is presented as a function of $\sqrt{U}F/\gamma^{3/2}$. Both the Gaussian result for the thermodynamic limit and numerical results with finite photon-densities are presented. At each of the bifurcation points, a cusp emerges in the thermodynamic limit. A maximum value close to 0.35 is predicted at the first bifurcation point.
The results for a finite nonlinearity $U$ exhibit a single peak around the first bifurcation point and a weak shoulder around the second one. 
Note that the entanglement increases in the thermodynamical limit ($U \to 0$ with $\sqrt{U} F$ fixed) proving that criticality enhances the quantum correlations as well.

\begin{figure}[h]
  \includegraphics[scale=0.65]{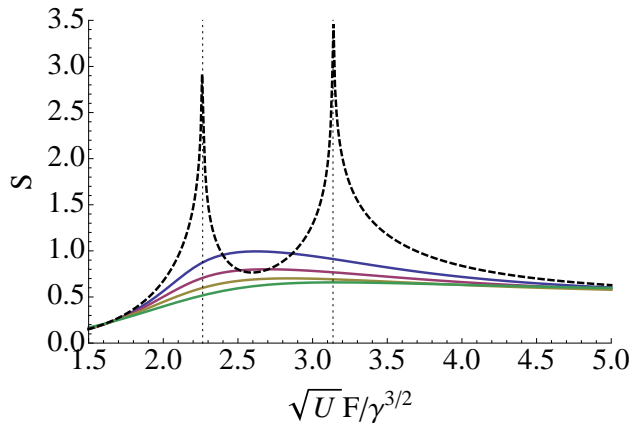}
  \caption{\label{Fig5} The von Neumann entropy $S$ as a function of $\sqrt{U}F/\gamma^{3/2}$. The dashed curve is the Gaussian result, which is valid in the thermodynamic limit ($U \rightarrow 0$ with $\sqrt{U} F$ fixed). The full curves correspond to finite values of the nonlinear interaction $U/\gamma = 0.1, 0.25, 0.5$ and $1$ (from top to bottom respectively). Other system parameters are the same as in Fig. \ref{Fig1}.}
\end{figure}

In Fig. \ref{Fig5} the von Neumann entropy $S$ is presented as a function of $\sqrt{U}F/\gamma^{3/2}$. Again, both the Gaussian prediction for the thermodynamic limit and the numerical results for a finite nonlinearity are presented. In the thermodynamic limit two narrow peaks are observed at the bifurcation points with a finite maximum. This indicates that the system becomes highly mixed at the bifurcation points. The results for a finite nonlinearity exhibit a single peak in the critical quantum regime. 
   
\section{Conlusions and perspectives \label{Sec5}}	
We have presented a theoretical investigation of the driven-dissipative Bose-Hubbard dimer in the regime where a phase transition with a spontaneous spatial symmetry breaking occurs. It was shown that a thermodynamic limit is well defined by letting the nonlinear interaction $U \rightarrow 0$ and the driving amplitude $F \rightarrow \infty$,  while keeping the product $\sqrt{U}F$ fixed. In such thermodynamic limit a second-order phase transition with spatial symmetry breaking is well defined.
We predict a large quantum entanglement of the mixed steady-state, which is maximized in the thermodynamical limit. Numerical solutions for finite excitation numbers confirm these results and show the finite-size deviations from the Gaussian theory.
Our results demonstrate that quantum entanglement occurs also in dissipative phase transitions and is enhanced by critical behavior. 
The behavior of entanglement and criticality in driven-dissipative Bose-Hubbard lattices (and other physical models) with many sites is an intriguing problem to explore in the future.

\acknowledgements{We gratefully acknowledge discussions with N. Bartolo, A. Bramati, R. Fazio, A. Yiacomotti, M. Hafezi, M. Labousse, J. Lolli, F. Minganti, A. Hu, S. Rodriguez and R. Rota. We acknowledge support from ERC (via the Consolidator Grant "CORPHO" No. 616233)}  

\bibliography{manusc}

\end{document}